\documentclass[lettersize,journal]{IEEEtran}

\usepackage{fancyhdr}
 
\usepackage{array}
\usepackage[caption=false,font=normalsize,labelfont=sf,textfont=sf]{subfig}
\usepackage{textcomp}
\usepackage{stfloats}
\usepackage{url}
\usepackage{verbatim}
\usepackage{graphicx}
 
\usepackage{cite}
\usepackage{amsmath,graphicx}
\usepackage{amssymb}
\usepackage{algpseudocode}
\usepackage{amsfonts}
\usepackage{fancyhdr}  %
\usepackage{cases}
\usepackage{extarrows}
\usepackage{algorithm}
\usepackage{multirow,tabularx}
\usepackage{mathtools}
\usepackage{xcolor}
\usepackage[english]{babel}
\usepackage{caption}
\usepackage{autobreak}
\usepackage{bbm}

\ifCLASSINFOpdf
\else
\fi
\begin{document} 

\title{    Multi-User Holographic MIMO Surfaces:\\ Channel Modeling and Spectral Efficiency Analysis }
\author{Li Wei, Chongwen Huang, George~C.~Alexandropoulos,~\IEEEmembership{Senior Member,~IEEE,}  Wei E. I. Sha,~\IEEEmembership{Senior Member,~IEEE,} Zhaoyang Zhang,~\IEEEmembership{Senior Member,~IEEE},  M\'{e}rouane~Debbah,~\IEEEmembership{Fellow,~IEEE} and Chau~Yuen,~\IEEEmembership{Fellow,~IEEE}
 
\thanks{L. Wei and C. Yuen are with the Engineering Product Development (EPD) Pillar, Singapore University of Technology and Design, Singapore 487372 (e-mails: wei\_li@mymail.sutd.edu.sg, yuenchau@sutd.edu.sg).}

\thanks{C.~Huang and Z.~Zhang are with College of Information Science and Electronic Engineering, Zhejiang University, Hangzhou 310027, China, and with International Joint Innovation Center, Zhejiang University, Haining 314400, China, and also with Zhejiang Provincial Key Laboratory of Info. Proc., Commun. \& Netw. (IPCAN), Hangzhou 310027, China (e-mails: \{chongwenhuang, ning\_ming\}@zju.edu.cn).}

\thanks{G.~C.~Alexandropoulos is with the Department of Informatics and Telecommunications, National and Kapodistrian University of Athens, Panepistimiopolis 	Ilissia, 15784 Athens, Greece. He also serves as a Principal Researcher at the Technology Innovation Institute, Abu Dhabi, United Arab Emirates.  (e-mail: alexandg@di.uoa.gr).}
 
\thanks{ Wei  E.  I.  Sha is with the College of Information Science and Electronic Engineering, Zhejiang University, Hangzhou 310027, China  (e-mail:  weisha@zju.edu.cn).  }
 
\thanks{ M. Debbah is with the Technology Innovation Institute, 9639 Masdar City, Abu Dhabi, United Arab Emirates (email: merouane.debbah@tii.ae) and also with CentraleSupelec, University Paris-Saclay, 91192 Gif-sur-Yvette, France.}
}

\maketitle

\thispagestyle{fancy}

\begin{abstract}
The multi-user Holographic Multiple-Input and Multiple-Output Surface (MU-HMIMOS) paradigm, which is capable of realizing large continuous apertures with minimal power consumption, has been recently considered as an energy-efficient solution for future wireless networks, offering increased flexibility in impacting electromagnetic (EM) wave propagation according to the desired communication, localization, and sensing objectives. The tractable channel modeling in MU-HMIMOS wireless systems is one of the most critical research challenges, mainly due to the coupling effect induced by the excessively large number of closely spaced patch antennas. In this paper, we focus on this challenge for the downlink of multi-user MIMO communications and extend an EM-compliant channel model to multi-user case, which is expressed in the wavenumber domain using the Fourier plane wave approximation. Based on the presented channel model, we investigate the spectral efficiency of maximum-ratio transmission and Zero-Forcing (ZF) precoding schemes. We also introduce a novel hardware efficient ZF precoder, leveraging Neumann series (NS) expansion to replace the required matrix inversion operation, which is very hard to be computed in the conventional way due to the extremely large number of patch antennas in the envisioned MU-HMIMOS communication systems. In comparison with the conventional independent and identical Rayleigh fading channels that ignore antenna coupling effects, the proposed EM-compliant channel model captures the mutual couplings induced by the very small antenna spacing. Our extensive performance evaluation results demonstrate that our theoretical performance expressions approximate sufficiently well the simulated achievable spectral efficiency with the considered linear precoding schemes, even for the highly correlated cases, thus verifying the effectiveness and robustness of the presented analytical framework. In addition, it is verified that the proposed NS-based ZF precoder achieves similar performance to conventional ZF, while requiring lower hardware complexity, thus, providing a hardware efficient solution for practical design of MU-HMIMOS communications systems.
\end{abstract}


\section{Introduction}\label{sec:intro}
In recent years, the demands for ubiquitous wireless communications are rapidly growing due to the explosive development of mobile devices and multimedia applications (including extending reality) \cite{9374451}. To satisfy those needs and requirements, the exploitation of the ultimate limits of wireless communication approaches is mandatory, in conjunction with the introduction and optimization of novel technologies providing larger bandwidths and improved energy efficiency, e.g., TeraHertz (THz) \cite{9325920} and millimeter-Wave (mmWave) communications \cite{Akyildiz2018mag}, as well as extreme Multiple-Input Multiple-Output (MIMO) catering for massive users and throughput \cite{shlezinger2020dynamic_all,9475156}. However, more profound requirements for data rate and massive connections are lately put forward for 6-th Generation (6G) wireless communications. In addition, the large available bandwidth at higher frequencies does not come for free, imposing several algorithmic and hardware design challenges. 

In MIMO systems, a Base Station (BS) equipped with multiple antenna elements serves a single or multiple users via spatial multiplexing \cite{CoBF_2016,9462554,4392826,6678765,6489506,9124713,RISE6G_COMMAG}. Techniques falling into this category are capable of improving Spectral Efficiency (SE), by increasing the number of antennas and/or the users. The massive MIMO version, which has been recently adopted in the 5-th Generation (5G) New Radio (NR), utilizes very large antenna arrays at the BS to serve multiple single-antenna users with simple beamforming, yielding boosted rate performance \cite{6241389}. In fact, even when the number of BS antennas grows large, the overall power consumption is kept fixed SE improves \cite{6457363}. In \cite{8501941}, the authors showcased that the hardware distortion in massive MIMO has negligible impact on the achievable SE performance. In \cite{9386231}, a cooperative Non-Orthogonal Multiple Access (NOMA) based MIMO system model for 6G wireless communications was proposed, targeting the rate improvement of cell-edge users. The authors of \cite{9506868} analyzed the asymptotic performance of multi-user MIMO communications with different linear receivers. However, although numerous techniques for combating signal attenuation are available for massive MIMO, this technology is challenges in mobility scenarios and is accompanied by the antenna and Radio Frequency (RF) front-end scalability problem. 

Fortunately, the technologies of metasurface-based antennas and Reconfigurable Intelligent Surfaces (RISs) provide feasible and engaging research directions towards realizing highly flexible antennas at low cost \cite{alexandg_2021, RISE6G_COMMAG, yanglianghnu20, 9501003,wanghuimingSecure,8972400,9475160,WavePropTCCN}. Specifically, an RIS is usually comprised of a large number of hardware-efficient and nearly passive reflecting elements, each of which can alter the phase of the incoming signal, without requiring a dedicated power amplifier \cite{holobeamforming,huang2019reconfigurable,9119122,9140329}. Thus, the deployment of RISs enables the manipulation of electromagnetic (EM) waves, and brings benefit in low-power, energy-efficient, high-speed, massive-connectivity, and low-latency wireless communications \cite{husha_LIS2,liaskos2018new,qingqing2019towards,WavePropTCCN,9206044,9110849,9388935}. Due to its advantages, the research on RIS-assisted communications has gained remarkable research interest \cite{9201173,Marco2019,9180053,9090356,9366805}. In \cite{8941126}, novel passive beamforming and information transfer technologies to enhance primary communication were proposed. Furthermore, NOMA in RIS-assisted communications was studied in \cite{liunoma2020} as a cost-effective solution for boosting spectrum/energy efficiency.       

Inspired by the potential of RISs in 6G wireless communications \cite{CE_overview_2022}, the Holographic MIMO Surface (HMIMOS) concept aims at going beyond massive MIMO systems \cite{9136592, demir2021channel, 8437634,9765526 }. Specifically, an HMIMOS system incorporates densely packed sub-wavelength patch antennas to achieve programmable wireless environments \cite{RISE6G_COMMAG}, and it is verified to boost competitiveness in many fields, including NOMA, unmanned aerial vehicles, mmWaves, and multi-antenna systems \cite{9530717}. The work in \cite{9110848,9443506} provided a comprehensive explanation of HMIMOS communications in wavenumber domain with plane waves.  The work \cite{9136592} proved theoretically the flexibility of HMIMOS configuration and its advantages in improving SE through intelligent environment configuration. The authors in \cite{9145091} investigated the mutual coupling matrix \cite{Nossek, alexandg_ESPARs} in large RIS-assisted single-user communication systems, and showed that large RISs can achieve super-directivity.    Benefiting from these merits, HMIMOS can be considered for many applications, including the extension of coverage, wireless power transfer, and indoor positioning \cite{9136592}. However, the full exploitation of HMIMOS is still challenging due to many non-trivial technical issues.   

One of the main challenges of the HMIMOS technology is multi-user channel modeling, due to the spatially continuous aperture realized by holographic patch antennas \cite{9475156}. The traditional channel models are not applicable for various reasons, e.g., the coupling between antennas, their massive numbers, and the surface-to-surface transmission. Thus, the shift towards infinite antennas and operating frequencies (i.e., mmWave and THz) requires an efficient EM model, since traditional independent Rayleigh fading models, which are based on the assumption of far-field EM wave propagation, might not be applicable \cite{8437634, sanguinetti2021wavenumber,9765526}. {Typically, in traditional channel models, the correlation between antennas is assumed to be zero with spacing larger than or equal to $\lambda/2$, however, when the spacing is less than $\lambda/2$, this will have the strong coupling effect in practice. Under such conditions, the spatial correlation \cite{9110848,9724113,NakagamiTWC_2009,NakagamiTVT_2010} among densely packed antennas cannot be ignored, and a tractable stochastic modeling tool is required to characterize the channel.}  In \cite{9374451}, beam pattern and channel estimation schemes for holographic RIS-assisted THz massive MIMO systems were presented, however, the channel was modeled in the beam domain, considering only the line-of-sight path. In \cite{9110848}, a spatially correlated small-scale fading model for single-user HMIMOS communication systems was proposed, employing samples of the random EM field to represent the EM channel.   The research in \cite{9154219,PhysRevApplied.16.064042,9650519} investigated the number of channel spatial Degrees of Freedom (DoF) in isotropic scattering environments, considering spatially-constrained apertures, and it was proved that the spatial DoF are proportional to the surface area, which is distinct from traditional analysis methods. When considering Multi-User HMIMOS (MU-HMIMOS) communication systems, another challenge is that simple linear precoding schemes, such as Zero-Forcing (ZF), with continuous-apertures surfaces are impractical due to the large amount of patch antennas. 

{Motivated by the EM-compliant channel model of \cite{9110848,9765526,9724113} for single-user HMIMOS communication systems, this paper introduces a convenient channel model for MU-HMIMOS systems, and theoretically analyzes the SE performance of different linear precoding, which are designed exploiting features of the proposed model. Typically, the wireless channel empowered by an HMIMOS with infinite numbers of patch antennas is assumed to be on the continuous EM space in the optimal setting. To facilitate practical HMIMOS-based system designs, the continuous EM channel is sampled according to a specific placement and orientation of the patch antennas. Hence, the actual communication can be considered as a functional analysis problem depending only on geometric relationships, and the involved channel can be constructed as a combination of complete basis function sets of transmitter/receiver surfaces. In this manner, the ultimate performance metric for communications, namely the intrinsic capacity of the sampled space wireless channel, can be studied. Specifically, due to the fact that the EM channel can be characterized by complete basis functions, the SE can be theoretically analyzed using the variances of the sampled	transmitter/receiver surfaces.} In addition, the MU-HMIMOS concept involves massive patch antennas, thus, even simple linear precoding schemes (e.g., ZF and Minimum Mean-Square Error (MMSE)) are computationally heavy and costly, due to the expensive matrix inversion operation \cite{9027848,7248549,8332955,6572301,6554990,9115725}. To solve this problem, we present a novel low-hardware-complexity  ZF precoding scheme that is based on a Neumann Series (NS) expansion, which replaces the expensive matrix inversion operation while being similar in terms of achievable sum rate with conventional ZF. The main contributions of this paper are summarized as follows:
\begin{itemize}
	\item  {Building on the Fourier plane wave theory developed in \cite{9110848}, we extend the work to MU-HMIMOS communication systems. Specifically, the finite number of sampling points that carry channel information are collected to represent the spatially continuous channel, and the sampling is dependent on the spacing and the total amount of the surface's patch antennas.}
	\item  {We investigate the SE of Maximum Ratio Transmission (MRT) and ZF precoding schemes in  MU-HMIMOS communications, and derive the corresponding achievable rate capitalizing on the characteristics of the proposed EM-compliant channel model. } 
	\item  {We introduce a hardware-efficient ZF precoding scheme, leveraging NS expansion to replace the matrix inversion, as it is practically impossible to implement this operation due to the extremely large number of patch antennas  in the envisioned MU-HMIMOS communication systems.}
	\item We prove the validity of our theoretical capacity through simulation results and investigate the impact of spacing among the HMIMOS patch antennas, providing various insights on the surface's design. 
\end{itemize}      

The remainder of this paper is organized as follows. In Section \ref{sec:EM_channel}, the extended EM channel modeling for MU-HMIMOS communication systems is presented.  Section~\ref{sec:sum_rate} presents the analytical formulas for the SE with the considered precoding schemes and a low complexity NS-based ZF precoding scheme. Our simulation results are given in Section~\ref{sec:simu}, while the concluded remarks of the paper are drawn in Section~\ref{sec:conclusion}.
	
\textit{Notation}: Fonts $a$, $\mathbf{a}$, and $\mathbf{A}$ represent scalars, vectors, and matrices, respectively. $\mathbf{A}^T$, $\mathbf{A}^H$, $\mathbf{A}^{-1}$, $\mathbf{A^\dag}$, and $\|\mathbf{A}\|_F$ denote transpose, Hermitian (conjugate transpose), inverse, pseudo-inverse, and Frobenius norm of $ \mathbf{A} $, respectively. $\mathbf{A}_{i,j}$ or $[\mathbf{A}]_{i,j}$ represents $\mathbf{A}$'s $(i,j)$-th element, while $[\mathbf{A}]_{i,:}$ and $[\mathbf{A}]_{:,j}$ stand for its $i$-th row and $j$-th column, respectively. $|\cdot|$ and $(\cdot)^*$ denote the modulus and conjugate, respectively. $\text{tr}(\cdot)$ gives the trace of a matrix, $\mathbf{I}_n$ (with $n\geq2$) is the $n\times n$ identity matrix, and $\mathbf{1}_n$ is a column vector with all ones.   $\delta_{k,i}$ equals to $1$ when $k=i$ or $0$ when $k\neq i$, and $\mathbf{e}_n$ is the $n$-th element unit coordinate vector with $1$ in the $n$-th basis and $0$'s in each $n^\prime$-th basis $\forall$$n^\prime \neq n$. Finally, notation ${\rm diag}(\mathbf{a})$ represents a diagonal matrix with the entries of $\mathbf{a}$ on its main diagonal,  $\delta(\cdot)$ is the Dirac delta function, and $\odot$ is the Hadamard product. Table~I includes a list with the mist frequently-used parameters and variables throughout the paper.

\begin{table}[!t]
	\caption{ {Parameters and variables used in this paper.} }
	\centering
	\begin{tabular}{ |m{2.2cm}| m{5.3cm}| }
		\hline
		 {Parameters/variables} &  {Definitions} \\
		\hline
		 $N_s, N_r$ &   Number of patch antennas at the BS and each user, respectively. \\ 
		\hline
		 $L_{s,x}, L_{s,y}$ & Horizontal and vertical length of the transmit surface.\\
		\hline
		 $L_{r,x}, L_{r,y}$ &  Horizontal and vertical length of the receive surface.\\
		\hline
		 $n_s, n_r$ &  Number of sampling points at transmit surface and receive surface, respectively. \\  
		\hline
		 $\boldsymbol{k}, \mathbf{k}, k$ &  Transmit wave vector, receive wave vector, and wavenumber, respectively.  \\ 
		\hline
		 $\mathbf{H}, \mathbf{H}_a$ & Channel and the equivalent wave-domain channel, respectively.\\ 
		\hline
		 $\mathbf{a}_s({\boldsymbol{k}}, \mathbf{s}), \mathbf{a}_r({\mathbf{k}}, \mathbf{r})$ & Source and receive responses in wave domain, respectively.\\
		\hline
		 $A\left(k_{x}, k_{y}, \kappa_{x}, \kappa_{y}\right)$, $S \left(k_{x}, k_{y}, \kappa_{x}, \kappa_{y}\right)$ & Spectral factor and spectral density at the point $(k_{x}, k_{y}, \kappa_{x}, \kappa_{y})$ in the wave domain, respectively.\\
		\hline
		 $\sigma^{2}\left(\ell_{x},\ell_{y},m_x,m_y\right)$ &  Wave-domain channel variance of the sampling point $\left(\ell_{x},\ell_{y},m_x,m_y\right)$. \\
		\hline
		 $\mathbf{\Phi}, \mathbf{V}$ &  Phase and precoding matrices, respectively.\\
		\hline
		 $\tilde{\mathcal{R}}_{\mathrm{(MRT)}},\tilde{\mathcal{R}}_{\mathrm{(ZF)}}$ &  Theoretical capacity bound of the MRT and ZF precoding systems, respectively.\\
		\hline
	\end{tabular}
\end{table} 

\section{EM-Compliant Channel Modeling} \label{sec:EM_channel}
In this section, we capitalize on  {the approximated Fourier plane-wave series expansion, for the case of far-field EM wave propagation, and present an EM-compliant far-field channel model for multi-user HMIMOS communication systems.}
\subsection{System Model}\label{subsec:signal model}
\begin{figure} 
	\begin{center}
		\centerline{\includegraphics[width=0.45\textwidth]{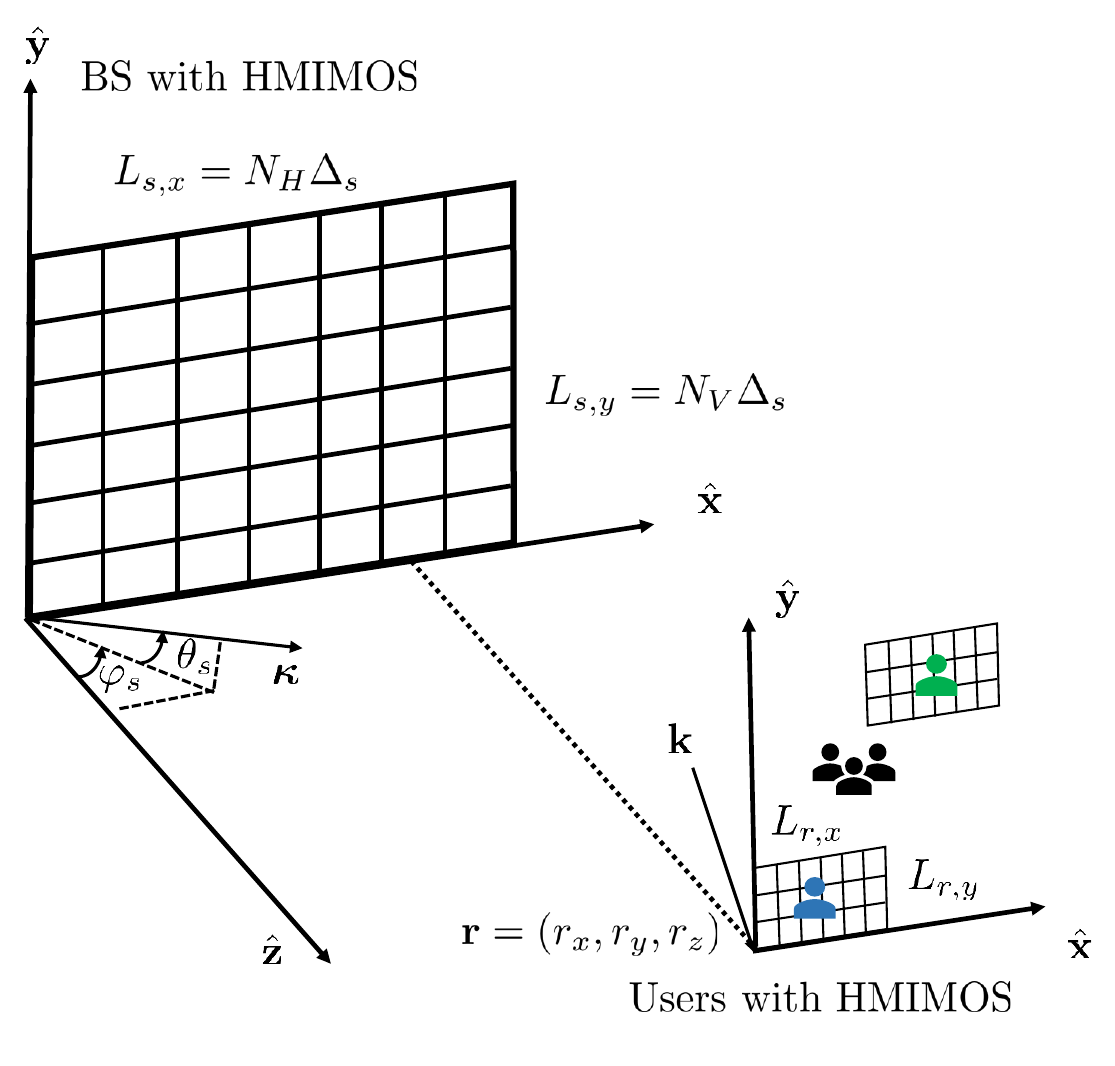}}  \vspace{-0mm}
		\caption{The considered MU-HMIMOS communication system consisting of an $N_s$ patch antennas at BS end and $M$ users with each having $N_r$ patch antennas.}
		\label{fig:Estimation_Scheme} 
	\end{center}
\end{figure}

Consider the downlink communication between a BS and a group of $M$ users that are both equipped with HMIMOS,  as shown in Fig$.$~\ref{fig:Estimation_Scheme}. The HMIMOS at BS side is comprised of $N_s=N_V N_H$  unit cells, each made from metamaterials that are capable of adjusting their reflection coefficients, and the patch antennas spacing $\Delta_s$ is below half of the wavelength $\lambda$. The horizontal and vertical length of HMIMOS is $L_{s,x}= N_H \Delta_s$ and $L_{s,y}=N_H \Delta_s$. The HMIMOS patch antennas at BS are indexed row-by-row by $n \in[1, N_s]$, thus the location of the $n$-th patch antenna with respect to the origin is   
\begin{equation}
	\mathbf{s}_{n}=[s_x,s_y,s_z]=\left[ 0,   i(n) \Delta_s,   j(n) \Delta_s\right]^{\mathrm{T}},
\end{equation} 
where $ i(n)=\bmod \left(n-1, N_H\right)$ and $ j(n)=\left\lfloor(n-1) / N_H \right\rfloor$ are the horizontal and vertical indices of element $n$, respectively. Notice that $\bmod(\cdot, \cdot)$ denotes the modulus operation and $\lfloor\cdot\rfloor$ is the truncation of argument. 

{The transmit vector at HMIMOS with the azimuth angle $\varphi$ and elevation angle $\theta$ is
	\begin{equation} \label{equ:transmit_vector}
		\begin{aligned}
			\mathbf{a}_s(\boldsymbol{\kappa}, \mathbf{s})&=\left[a_{s,1}(\boldsymbol{\kappa}, \mathbf{s}),\ldots,a_{s,N_s}(\boldsymbol{\kappa}, \mathbf{s})\right] \\
			&=\left[e^{\mathrm{j} \boldsymbol{\kappa}(\varphi, \theta)^{\mathrm{T}} \mathbf{s}_{1}}, \ldots, e^{\mathrm{j} \boldsymbol{\kappa}(\varphi, \theta)^{\mathrm{T}} \mathbf{s}_{N_s}}\right]^{\mathrm{T}},
		\end{aligned}
	\end{equation} 
	where $\boldsymbol{\kappa}(\varphi, \theta) \in \mathbb{R}^{3}$ is the transmit wave vector  
	\begin{equation}
		\begin{aligned}
			\boldsymbol{\kappa}(\varphi, \theta)&=[\kappa_{x},\kappa_{y},\kappa_{z}]\\
			&=k[\cos (\theta) \cos (\varphi),  \cos (\theta) \sin (\varphi), \sin (\theta)]^{\mathrm{T}},
		\end{aligned}
	\end{equation}
	with $k=\frac{2 \pi}{\lambda}$ being wavenumber.}

	Similarly, each user is equipped with $N_r$ patch antennas with the spacing $\Delta_r$. $L_{r,x}$ and $L_{r,y}$ denote the horizontal and vertical length, respectively. We assume that there are $M$ users, and the location of $m$-th user is denoted by  $\mathbf{r}_m=[r_{m,x},r_{m,y},r_{m,z}], m=1,\ldots,M$. The receive vector is $\mathbf{a}_r(\mathbf{k}, \mathbf{r})=\left[a_{r,1}(\mathbf{k}, \mathbf{r}),\ldots,a_{r,M}(\mathbf{k}, \mathbf{r})\right]^{\mathrm{T}} \in \mathbb{C}^{M\times 1}$, with 
	\begin{equation} \label{equ:receive_vector}
		\begin{aligned}
			\mathbf{a}_{r,m}(\mathbf{k}, \mathbf{r}) =\left[e^{\mathrm{j} \boldsymbol{k} ^{\mathrm{T}} \mathbf{r}^{(m)}_{1}}, \ldots, e^{\mathrm{j} \boldsymbol{k} ^{\mathrm{T}} \mathbf{r}^{(m)}_{N_r}}\right]^{\mathrm{T}}, m=1,\ldots,M,
		\end{aligned}
	\end{equation} 
	where $\boldsymbol{k}  \in \mathbb{R}^{3}$ is the receive wave vector $\boldsymbol{\kappa}=[k_{x},k_{y},k_{z}]$. 

\subsection{Channel Modeling for Individual Users} 
Similar to Fourier transformation between time domain and frequency domain, there also exists transformation between the space domain and wavenumber domain. Specifically, the $(m,n)$-entry of the spatial domain channel $\mathbf{H} \in \mathbb{C}^{N_r \times N_s}$ can be obtained by the wavenumber domain channel $\mathbf{H}_a\left(k_{x}, k_{y}, \kappa_{x}, \kappa_{y}\right)$ [54, Eq. 4], i.e., 
\begin{equation} \label{equ:channel_wavenumber}
	\begin{aligned}
		[\mathbf{H}]_{mn}&\!=\!\frac{1}{(2 \pi)^{2}} \iiiint_{\mathcal{D} \times \mathcal{D}} \! {a}_{r,m}(\mathbf{k}, \mathbf{r}) H_{a}\!\left(k_{x}, k_{y}, \kappa_{x}, \kappa_{y}\right) \\
		&\qquad  \qquad \qquad {a}_{s,n}(\boldsymbol{\kappa}, \mathbf{s}) d k_{x} d k_{y} d \kappa_{x} d \kappa_{y},
	\end{aligned}
\end{equation}
where ${a}_{s,n}(\boldsymbol{\kappa}, \mathbf{s})$ is the $n$-th element in \eqref{equ:transmit_vector}, ${a}_{r,m}(\mathbf{k}, \mathbf{r})$ is the $m$-th element in \eqref{equ:receive_vector}, and integration region is $\mathcal{D}=\{(k_x,k_y) \in \mathbb{R}^2: k_x^2+k_y^2 \leq \kappa ^2 \}$ since the wave propagates along $z$-direction, and projection lies in X-Y plane.

The equivalent wavenumber domain channel is [54, Eq. 9]
\begin{equation} 
	\begin{aligned}
		&H_{a}\!\left(\!k_{x}, k_{y}, \!\kappa_{x},\! \kappa_{y}\!\right)\!=\!S^{1/2}\!\left(k_{x},\! k_{y},\! \kappa_{x},\! \kappa_{y}\right) \!W\!\left(k_{x},\! k_{y}, \! \kappa_{x}, \! \kappa_{y}\right)\\
		&=\frac{\kappa \eta}{2} \frac{  A\left(k_{x}, k_{y}, \kappa_{x}, \kappa_{y}\right) W\left(k_{x}, k_{y}, \kappa_{x}, \kappa_{y}\right)}{k_{z}^{1/2}\left(k_{x}, k_{y}\right) \kappa_{z}^{1/2} \left(\kappa_{x}, \kappa_{y}\right)},
	\end{aligned}
\end{equation} 
where the spectral density is $S \left(k_{x}, k_{y}, \kappa_{x}, \kappa_{y}\right)=\frac{  A^2\left(k_{x}, k_{y}, \kappa_{x}, \kappa_{y}\right) }{k_{z} \left(k_{x}, k_{y}\right) \kappa_{z}  \left(\kappa_{x}, \kappa_{y}\right)}$, and the spectral factor $A\left(k_{x}, k_{y}, \kappa_{x}, \kappa_{y}\right)$ is an arbitrary real-valued, non-negative function that is dependent on scattering environment  [54, Theorem 1]. In isotropic scattering environment, $A\left(k_{x}, k_{y}, k_{z}\right)=\frac{2 \pi} {\sqrt{k}}$ with unit channel power \cite{9110848}. $W\left(k_{x}, k_{y}, \kappa_{x}, \kappa_{y}\right)\sim \mathcal{CN}(0,1)$ is included to represent random characteristics. 

In fact, $H_{a}\left(k_{x}, k_{y}, \kappa_{x}, \kappa_{y}\right)$ is the intermediate matrix between the transmit vector and receive vector in angular domain, which depends on the scattering environment and array geometry. Moreover, the more eigenvalues $H_{a}\left(k_{x}, k_{y}, \kappa_{x}, \kappa_{y}\right)$ has, the more strongly connected channels exist.

In wavenumber domain, the equivalent angular channel is a sparse and there are only finite non-zero elements, thus, the \eqref{equ:channel_wavenumber} can be approximated with finite sampling points in the effective bandwidth, i.e., the Fourier plane wave series expansion is non-zero only within the lattice ellipse \cite{9724113}
\begin{equation}
	\begin{aligned}
		&\mathcal{E}_{s}\!=\!\left\{\!\left(m_{x}, m_{y}\right) \in \mathbb{Z}^{2}\!:\!\left(\!m_{x} \lambda / L_{s,x}\!\right)^{2}\!+\!\left(m_{y} \lambda / L_{s,y}\right)^{2} \!\leq\! 1\!\right\}\!, \\
		&\mathcal{E}_{r}\!=\!\left\{\left(\ell_{x}, \ell_{y}\right) \in \mathbb{Z}^{2}\!:\left(\ell_{x} \lambda / L_{r, x}\right)^{2}+\left(\ell_{y} \lambda / L_{r, y}\right)^{2} \!\leq 1\!\right\},
	\end{aligned}
\end{equation}
at the source and receiver, respectively. The cardinalities of the sets $\mathcal{E}_{s}$ and $\mathcal{E}_{r}$ are $n_{s}=\left|\mathcal{E}_{s}\right|$ and $n_{r}=\left|\mathcal{E}_{r}\right|$, respectively. To fully exploiting the channel information, the numbers of patch antennas at BS and each user are $N_s\geq n_s$ and $N_r \geq  n_r$, respectively \cite{9724113, franceschetti2017wave}.

\begin{figure} 
	\begin{center}
		\centerline{\includegraphics[width=0.5\textwidth]{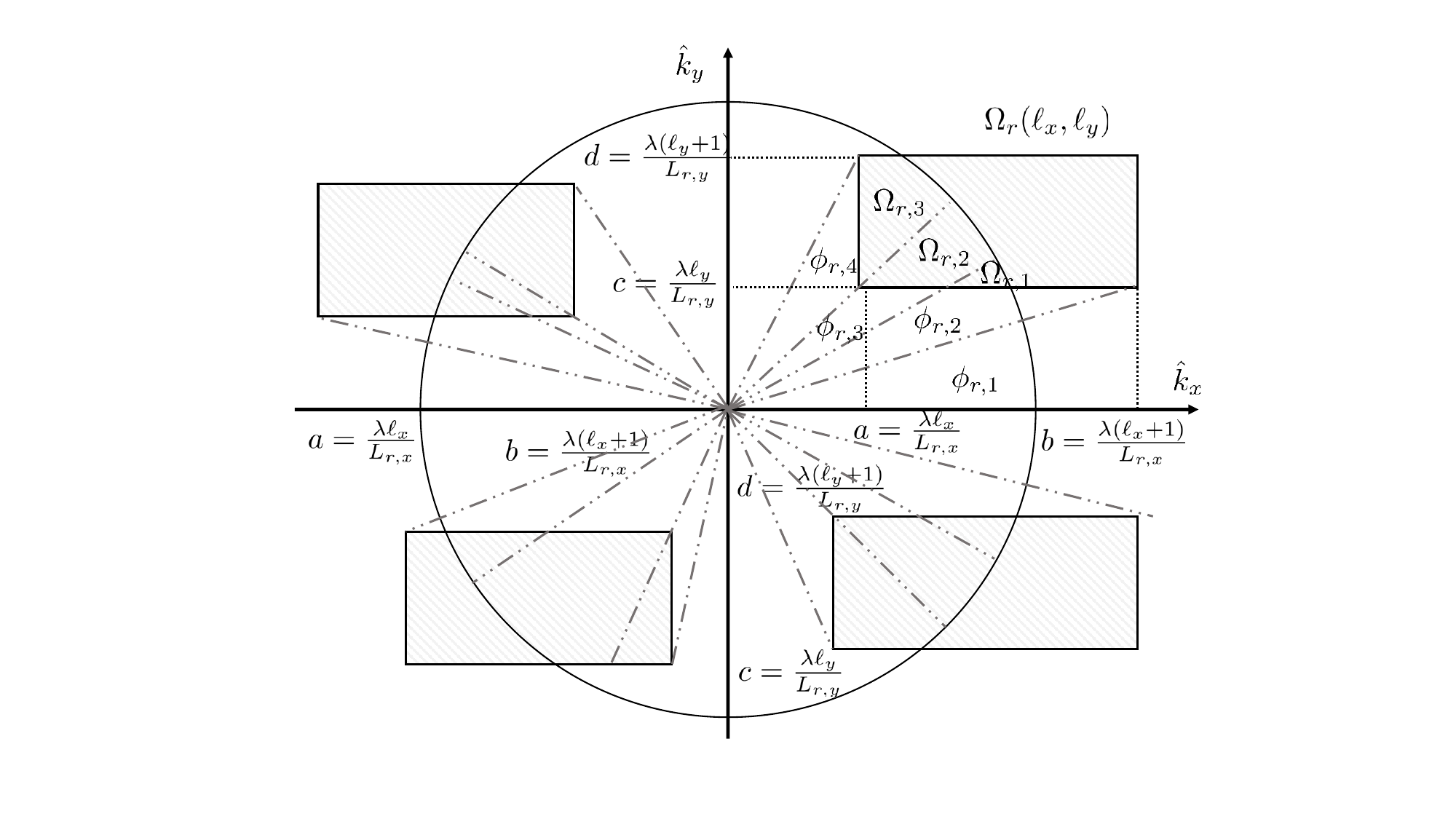}}  \vspace{-0mm}
		\caption{  Integration regions $\Omega_{r}(\ell_{x},\ell_{y})$ of the variances $\sigma^{2}\left(\ell_{x},\ell_{y},m_x,m_y\right)$  in four orthants ($L_{r,x}\leq L_{r,y}$).}
		\label{fig:IntegOrthants}  
	\end{center}
\end{figure}

Combining wavenumber domain channel at $n_s$ and $n_r$ sampling points in transmit and receive surface to approximate spatial channel $\mathbf{H}$, we have [54, Theorem 2]  
\begin{equation}
\begin{aligned}
	[\mathbf{H}]_{mn} \approx &\sum_{\left(\ell_{x}, \ell_{y}\right) \in \mathcal{E}_{r}} \sum_{\left(m_{x}, m_{y}\right) \in \mathcal{E}_{s}}  H_{a}\left(\ell_{x}, \ell_{y}, m_{x}, m_{y}\right)\\
	&\qquad  {a_{r,m}\left(\ell_{x}, \ell_{y}, \mathbf{r}\right) a_{s,n}\left(m_{x}, m_{y}, \mathbf{s}\right)},
\end{aligned}	 
\end{equation}
where the Fourier coefficient is  
\begin{equation}
	H_{a}\left(\ell_{x}, \ell_{y}, m_{x}, m_{y}\right) \sim \mathcal{CN} \left(0, \sigma^{2}\left(\ell_{x}, \ell_{y}, m_{x}, m_{y}\right)\right),
\end{equation}
with the variance $\sigma^{2}\left(\ell_{x}, \ell_{y}, m_{x}, m_{y}\right)$ is the variance of sampling point $\left(\ell_{x}, \ell_{y}, m_{x}, m_{y}\right)$, and it is separable under scattering separability, which can be computed from  \cite{9110848}
\begin{equation} \label{equ:variance_polar_sub}
	\begin{aligned}
		&\sigma^{2}\left(\!\ell_{x},\!\ell_{y},\!m_x,\!m_y\!\right)  \propto   \iiiint_{\hat{\mathcal{S}}_{s} \times \hat{\mathcal{S}}_{r} } \!\! \! \! 	 \mathbbm{1}_{\hat{\mathcal{D}}}\left(k_{x}, \! k_{y}\right)  \! \mathbbm{1}_{\hat{\mathcal{D}}} \!\left( \!\kappa_{x}, \! \kappa_{y} \!\right)  \\
		&\qquad \frac{  A^2\left(\hat k_{x}, \hat k_{y}, \hat \kappa_{x}, \hat \kappa_{y}\right) }{\hat k_{z}  \!\left( \!\hat k_{x}, \! \hat k_{y} \!\right) \hat \kappa_{z}  \! \left( \!\hat \kappa_{x}, \! \hat \kappa_{y} \!\right)}  d \hat k_{x} d \hat k_{y} d \hat \kappa_{x} d \hat \kappa_{y} \\
		&=\!\sum_{i=1}^{i=3} \!\sum_{j=1}^{j=3}  \!  \iiiint_{\Omega_{s,j}\left(m_{x},\! m_{y}\right) \times \Omega_{r,i}\left(\ell_{x}, \!\ell_{y}\right)}  \! \!  \!  \! \!\!\!\! \!\!\!\!  \!\!\!\! \!\!\!\! \!  \!  \!\!\!\! \!\!\!\! A^2\left(\theta_r,   \phi_r, \theta_s,   \phi_s\right)    d \Omega_{r } d \Omega_{s },
	\end{aligned}
\end{equation}
where  $(\hat k_x,\hat k_y, \hat k_z)= (k_x,k_y,k_z)/k$ are normalized wavevector coordinates, and integration region in receive wavenumber domain $\Omega_{r}(\ell_{x},\ell_{y})$ is divided into three subregions as shown in Fig.~\ref{fig:IntegOrthants}, $\Omega_{s}(m_{x},m_{y})$ is divided in the same way. The details refer to [48, Appendix IV.C]  

Collecting all variances in $\mathbf{\Sigma}^{(m)}=  (\boldsymbol{\sigma}_r^{(m)}  \mathbf{1}_{n_s}^{T}) \odot (\mathbf{1}_{n_r}    \boldsymbol{\sigma}_s^{T})$,   where $\boldsymbol{\sigma}_r^{(m)} \in \mathbb{R}^{n_r \times 1}$ and $\boldsymbol{\sigma}_s \in \mathbb{R}^{n_s \times 1}$ collect $\{\sqrt{N_r} \sigma_r^{(m)}(\ell_x,\ell_y)\}$ and $\{\sqrt{N_s} \sigma_s(m_x,m_y)\}$, respectively [54, Corollary 2]. ${\sigma}_{r,k}^{(m)}$ is the $k$-th element of $\boldsymbol{\sigma}_{r}^{(m)}$, and ${\sigma}_{s,i}$ is the $i$-th element of $\boldsymbol{\sigma}_{s}$, these variances are applied to capacity analysis in the next section.

\subsection{Channel Modeling for Multiple Users}\label{sec:iso_channel}
This part extends the individual user channel modeling from the previous subsection to the multi-user case, we assume that different users are independently distributed in space, thus, the multi-user involved channel matrix can be decomposed into multiple channel matrices that corresponds to different users. For simplicity, the isotropic scattering environment is considered. To fully characterize EM channel, the HMIMOS patch antennas $N_s \geq \frac{4}{\lambda^{2}} L_{s,x} L_{s,y} \geq n_s$ at the BS end, and each user is equipped with $N_r \geq \frac{4}{\lambda^{2}} L_{r, x} L_{r, y} \geq n_r$ patch antennas, where $\Delta_s$ and $\Delta_r$ are patch antennas spacing at BS and user end. The channel of $m$-th user in matrix form $\mathbf{H}^{(m)} \in \mathbb{C}^{N_r \times N_s}$ is \cite{9724113}
\begin{equation}
\begin{aligned}
	&\mathbf{H}^{(m)} \! = \!\sqrt{N_r \! N_s} \! \sum_{\left(\!\ell_{x}, \ell_{y}\!\right) \!\in \mathcal{E}_{\!r}} \sum_{\left(\!m_{x}, m_{y}\!\right) \!\in \!\mathcal{E}_{\!s}} H_{a}^{(\!m\!)}\left(\ell_{x},\! \ell_{y},\! m_{x},\! m_{y}\right) \\
	& \qquad\qquad {\mathbf{a}_{r}\left(\ell_{x}, \ell_{y}, \mathbf{r}^{(m)}\right) \mathbf{a}_{s}^{H}\left(m_{x}, m_{y}, \mathbf{s}\right)} ,
\end{aligned}	
\end{equation}
where the Fourier coefficient 
\begin{equation}
	H_{a}^{(m)}\left(\ell_{x}, \!\ell_{y},\! m_{x},\! m_{y}\right) \!\sim\! \mathcal{N}_{\mathbb{C}}\left(0, \sigma^{2}_{(m)}\left(\ell_{x}, \!\ell_{y}, \!m_{x},\! m_{y}\right)\right),
\end{equation}
and  \cite{9724113}
\begin{equation}
	\begin{aligned}
		&[a_{s}\left(m_{x},\! m_{y}, \!\mathbf{s}\right)]_j\!= \!\frac{1}{\!\sqrt{N_s}\!} e^{\!-\!\mathrm{j}\!\left(\!\frac{2 \!\pi\!}{L_{\!s\!,\!x}} \!m_{x} \!s_{\!x_j}\!+\!\frac{2 \!\pi}{L_{\!s\!,\!y}} \!m_{\!y}\! s_{\!y_j\!}\!+\!\gamma_{\!s}\!\left(\!m_{\!x}\!, \!m_{\!y}\right)\! s_{z_j}\!\right)\!}\!,\\
		&\qquad \qquad \qquad  j=1,\ldots, N_s, \\
		&[a_{r}\!\left(\ell_{x}\!, \!\ell_{y}, \!\mathbf{r}^{(m)}\!\right)\!]_i\!=\! \frac{1}{\!\sqrt{N_r}\!} e^{\!\mathrm{j}\!\left(\!\frac{2 \!\pi}{L_{r,\! x}} \!\ell_{x}\! r_{x_i}^{(m)}\!+\!\frac{2 \!\pi}{\!L_{r, \!y}} \!\ell_{y}\! r_{y_i}^{\!(m)\!}\!+\!\gamma_{\!r}\!\left(\!\ell_{x},\! \ell_{y}\!\right) \!r_{z_i}^{(\!m\!)}\right)\!}\!,\\
		& \qquad \qquad \qquad  i=1,\ldots, N_r.
	\end{aligned}
\end{equation}

Denote $\mathbf{U}_s \in \mathbb{C}^{N_s \times n_s}$ as the matrix collecting the all variances of $n_s$ sampling points in $a_{s}\left(m_{x},\! m_{y}, \!\mathbf{s}\right)$, and  $\mathbf{U}_r^{(m)} \in \mathbb{C}^{N_r \times n_r}$ as the matrix collecting the all variances of $n_r$ sampling points in $a_{r}\!\left(\ell_{x}\!, \!\ell_{y}, \!\mathbf{r}^{(m)}\!\right)$. Specifically, $\mathbf{U}_s  \mathbf{U}_s^{H}=\mathbf{I}_{n_s}$ and $\mathbf{U}_r^{(m)}  \mathbf{U}_r^{{(m)} H}=\mathbf{I}_{n_r}$ since their columns describe discretized transmit and receive plane-wave harmonics  [54, Eq. 46].

Based on the above observation, the channel matrix can be approximated by spatial sampling $n_s$ and $n_r$ points at $j=1,\ldots,N_s$-th patch antenna and the $i=1,\ldots,N_r$-th received antenna for $m$-th user. Thus, \cite{9724113}
\begin{equation}
	\begin{aligned}
		\mathbf{H}^{(m)}=\mathbf{U}_r^{(m)} \mathbf{H}_a^{(m)}  \mathbf{U}_s^{H}=\mathbf{U}_r^{(m)} \left(\mathbf{\Sigma}^{(m)} \odot \mathbf{W}\right)  \mathbf{U}_s^{H},
	\end{aligned}
\end{equation}
where $\mathbf{H}_a^{(m)} = \mathbf{\Sigma}^{(m)} \odot \mathbf{W} \in \mathbb{C}^{n_r\times n_s}$ collects  $\sqrt{N_r N_s} 	H_{a}^{(m)}\left(\ell_{x}, \ell_{y}, m_{x}, m_{y}\right)$ for all $n_r n_s$ sampling points,  and $\mathbf{W}\sim \mathcal{CN}(0,\mathbf{I}_{n_r n_s})$.

Thus, the correlation matrix of one user takes the form   \cite{9724113}
\begin{equation}
	\begin{aligned}
		&\mathbf{R}_{\mathbf{H} }^{(m)}\!=\! \mathbb{E}\left\{ \operatorname{vec} (\mathbf{H}^{(m)}) \operatorname{vec} (\mathbf{H}^{(m)}) ^{H}\right\} =\left(\mathbf{U}_{s}^{*} \!\otimes\! \mathbf{U}_{r}^{(m)}\right)\\
		& \mathbb{E}\left\{\!\left(\!\operatorname{vec}\left(\mathbf{\Sigma}^{(m)}\right) \odot \operatorname{vec}\left(\mathbf{W}\!\right) \!\right) \left(\operatorname{vec}\left(\mathbf{W}\right)  \! \odot \! \left(\mathbf{\Sigma}^{(m)}\!\right)\!\right)^{H} \!\right\}\\
		&\left(\mathbf{U}_{s}^{\mathrm{T}} \otimes \mathbf{U}_{r}^{{(m)} H}\right).
	\end{aligned}
\end{equation}

The channel $\mathbf{H}\in\mathbb{C}^{N_r M \times N_s}$ is given by 
\begin{equation}
	\begin{aligned}
		\mathbf{H}		&= \left[ \mathbf{U}_s^{*} {\mathbf{H}_a^{(1)}}^{T} {\mathbf{U}_r^{(1)}}^{T} ,\ldots,\mathbf{U}_s^{*} {\mathbf{H}_a^{(M)}}^{T} {\mathbf{U}_r^{(M)}}^{T}  \right]^{T} \\
		&=\mathbf{U}_r \mathbf{H}_a   \mathbf{U}_s^{H},
	\end{aligned}
\end{equation}
where $\mathbf{U}_r $ incorporates all received responses for $M$ users, $\mathbf{U}_s $ incorporates all transmit responses for all users, and $\mathbf{H}_a=\left[{\mathbf{H}_a^{(1)}}^{T},\ldots,{\mathbf{H}_a^{(M)}}^{T}  \right]^{T} $ contains all equivalent channels in wavenumber domain for all users.

The correlation matrix is 
\begin{equation}
	\begin{aligned}
		\mathbf{R}_{\mathbf{H}}&=\mathbb{E} \{ \operatorname{vec}(\mathbf{H})   \operatorname{vec}(\mathbf{H})^{\mathrm{H}} \} \\
		&= 
		\left[\begin{array}{llll}
			\mathbf{R}_{\mathbf{H}^{(1)}} & 0 & \cdots & 0 \\
			0 & \mathbf{R}_{\mathbf{H}^{(2)}} & \cdots & 0\\
			0 & 0 & \ddots & 0 \\ 
			0 & 0 & \cdots & \mathbf{R}_{\mathbf{H}^{(M)}} 
		\end{array}\right].
	\end{aligned}
\end{equation}
 
It can be observed from the above equations that the complex spatially correlated channel matrix $\mathbf{H}^{(m)}\in \mathbb{C}^{N_r \times N_s}$ is equivalent to a low-dimensional channel matrix $\mathbf{H}_a^{(m)}\in \mathbb{C}^{n_r \times n_s}$ since $N_s\geq n_s$ and $N_r \geq n_r$.
 
\section{Achievable Rate with Linear Precoding} \label{sec:sum_rate}
In this section, we derive the SE of the downlink MU-HMIMOS communication system adopting linear detectors (MRT and ZF). The perfect CSI is obtained at the receiver, and the phase matrix is assumed to be perfectly configured. Different from the conventional precoding analysis of MIMO system, the MU-HMIMOS system is equipped with a large number of patch antennas at both transmitter and receiver, which requires excessively high operational cost in practice, especially in the matrix inversion operation. In addition, in traditional MIMO system, each element in channel matrix is normally assumed to have the same unit variance, however, each entry in the constructed EM channel has different variances. Thus, the traditional analysis methods in large MIMO system cannot be directly applied in the considered scenario.  Based upon the equivalent wavenumber domain channel, we present a novel capacity evaluation method that requires the less computational cost and a hardware efficient  NS-based ZF precoding scheme.

\subsection{Ergodic Rate}
The HMIMOS at the transmitter is composed of $N_s$ patch antennas, and each user is equipped with $N_r$ patch antennas. There are total $M$ users, thus, the downlink MU-HMIMOS wireless communication system model is
\begin{equation}
\begin{aligned}
	\mathbf{y}&=\sqrt{p_u} \mathbf{H}   \mathbf{\Phi} \mathbf{V} \mathbf{x}+\mathbf{w}=\sqrt{p_u} \mathbf{U}_r  \mathbf{H}_a \mathbf{U}_s^{H} \mathbf{\Phi} \mathbf{V}  \mathbf{x} +\mathbf{w},
\end{aligned}
\end{equation}
where $\mathbf{y}\in \mathbb{C}^{N_r M \times 1}$ is the received signal; $\mathbf{H}\in \mathbb{C}^{N_r M \times N_s}$ is the channel matrix; $\mathbf{\Phi}=\operatorname{diag} (\boldsymbol{\phi})\in \mathbb{C}^{N_s \times N_s}$, $\boldsymbol{\phi}=[e^{j\phi_1},\ldots,e^{j\phi_{N_s}}]^{T}\in \mathbb{C}^{N_s\times 1}$ is the phase vector of HMIMOS part composed of $N_s$ patch antennas.  $ {p_u}$ is the transmitted power;  $\mathbf{V}\in \mathbb{C}^{N_s \times N_r M}$ is the precoding matrix; $\mathbf{x} \in \mathbb{C}^{ N_r M \times 1}$ is the transmitted signal, and $\mathbf{w}\in \mathbb{C}^{N_r M \times 1}$ is additive Gaussian noise that has i.i.d. elements with zero mean and variance $\sigma_w^2$. 

Let $\mathbf{y}_a= \mathbf{U}_r^{H} \mathbf{y}\in \mathbb{C}^{n_r M \times 1}$ the received signal in wavenumber domain, $\tilde{\mathbf{H}}_a=\mathbf{H}_a \mathbf{U}_s^{H} \mathbf{\Phi} \in \mathbb{C}^{ n_r M\times N_s}$ is the equivalent channel that incorporates the phase matrix. Thus, 
\begin{equation}
		\mathbf{y}_a =\sqrt{p_u} \tilde{\mathbf{H}}_a    \mathbf{V} \mathbf{x}+\mathbf{w}.
\end{equation}

The final signal received by $m$-th user at $i$-th received sampling point  can be given by
\begin{equation}
\begin{aligned}
	[{y}^{(m)}_a]_i=&[\tilde{\mathbf{H}}^{(m)}_a]_{i,:}   [\mathbf{V}^{(m)} ]_{:,i}   {x}^{(m)}_{i} \\
	&+  [\tilde{\mathbf{H}}^{(m)}_a]_{i,:}    \sum_{i'\neq i}^{n_r}   [\mathbf{V}^{(m)} ]_{:,i'}  {x}^{(m)}_{i}   +  {w}^{(m)}_{i},
\end{aligned}	
\end{equation}
where $\mathbf{x}^{(m)}=[x^{(m)}_{1} , \ldots, x^{(m)}_{N_r} ]\in \mathbb{C}^{N_r \times 1}$ is the transmitted signal of $m$-th user, with $x^{(m)}_{i}$ the $i$-th entry.  $ \mathbf{V}^{(m)}\in \mathbb{C}^{N_s \times N_r}$ is the $m$-th sub-block in precoding matrix corresponding to the $m$-th user, and $ [\mathbf{V}^{(m)} ]_{:,i} $ is the $i$-th column. $\tilde{\mathbf{H}}^{(m)}_a$ is the $m$-th sub-block of the channel channel matrix $\tilde{\mathbf{H}}_a $, and $[\tilde{\mathbf{H}}^{(m)}_a]_{i,:} $ is the $i$-th row.  $\mathbf{y}^{(m)} \in \mathbb{C}^{N_r \times 1}$ is the received signal of $m$-th user, and $[{y}^{(m)}_a]_i$ is the $i$-th element. Therefore, the achievable rate of the $m$-th user at $i$-th received response point $i=1,\ldots,n_r$ can be given by
\begin{equation}
	\begin{aligned}
		\mathcal{R}^{(\!m\!), i}\!=\!\mathbb{E}\! \left \{ \!\log_2\!\left(\!1\!\!+\!\!\frac{ p_u \left|[\tilde{\mathbf{H}}^{(m)}_a]_{i,:}   [\mathbf{V}^{(m)} ]_{:,i}   \right|^2}  {p_u\! \left|\!  [ \tilde{\mathbf{H}}^{(\!m\!)}_a\!]_{i,:}  \!  \sum_{\!i'\neq i}^{\!n_r}  \! [\mathbf{V}^{(\!m\!)} \!]_{: \!, i'} \!  \right|^2\! \!\!+\! \sigma_w^2}\!\right) \! \!\right\}\!.\!
	\end{aligned}
\end{equation}

\subsection{MRT Precoding}
In MRT precoding, $\mathbf{V}=\alpha_{\rm MRT}  \tilde{\mathbf{H}}_a ^{H}$, where $\alpha_{\rm MRT}$ is the normalization coefficient to satisfy the power constraint  $\mathbb{E}\{ \operatorname{Tr} ( \mathbf{V}    \mathbf{V} ^{H}  )\}=1$, we have \begin{equation}
	\begin{aligned}
		\alpha_{\rm MRT}& = \sqrt{\frac{1} {\mathbb{E}\{ \operatorname{Tr} (\tilde{\mathbf{H}}_a \tilde{\mathbf{H}}_a ^{H}  )\} } } = \sqrt{\frac{1}{  \mathbb{E}\{ \operatorname{Tr} (\mathbf{H}_a  \mathbf{H}_a^{H} )\} } } \\
		&= \sqrt{\frac{1}{\sum_{i=1}^{n_r} \sum_{j=1}^{n_s}  {\sigma_{r,i}^2\sigma_{s,j}^2  } } }  .
	\end{aligned}
\end{equation}

This is equivalent to normalize each column of precoding matrix $\mathbf{V}$, i.e.,   $\mathbf{V}_{:,i}=\frac{ \tilde{\mathbf{g}}_i}{ \| \mathbf{V}  \|_{F} }$, where $\tilde{\mathbf{g}}_i$ is the $i$-th row of equivalent channel $\tilde{\mathbf{H}}_a$, i.e., $\tilde{\mathbf{H}}_a^{H} =[\tilde{\mathbf{g}}_1, \ldots, \tilde{\mathbf{g}}_{n_r} ]$,  and $\alpha_{\rm MRT}=\frac{1}{{ \| \mathbf{V}  \|_{F} }}$.   Thus, 

\begin{equation} \label{sm_mrc_es}
	\begin{aligned}
		&\mathcal{R}^{(m),i}_{({\rm MRT})} \\
		&  =\mathbb{E}\! \left \{\! \log_2\!\left(\!1\!+\!\frac{p_u \alpha_{\rm MRT}^2 \left|  [\tilde{\mathbf{H}}_a^{(m)} [\tilde{\mathbf{H}}_a^{(m)}]^{H}]_{i,i} \right|^2 }  {p_u \alpha_{\rm MRT}^2 \left| [\tilde{\mathbf{H}}^{(m)}_a]_{i,:}    \sum_{i'\neq i}^{n_r}   [\tilde{\mathbf{H}}_a^{(m)} ]_{:,i'}   \right|^2\! +\! \sigma^2_w}\!\right) \! \right\} \\
		&=\mathbb{E}\! \left \{\! \log_2\!\left(\!1\!+\!\frac{p_u \alpha_{\rm MRT}^2 \left|  [{\mathbf{H}}_a^{(m)} [{\mathbf{H}}_a^{(m)}]^{H}]_{i,i} \right|^2 }  {p_u \alpha_{\rm MRT}^2 \left| [{\mathbf{H}}^{(m)}_a]_{i,:}    \sum_{i'\neq i}^{n_r}   [{\mathbf{H}}_a^{(m)} ]_{:,i'}   \right|^2\! +\! \sigma^2_w}\!\right) \! \right\} \\
		& {=} \mathbb{E} \left \{ \log_2\left(1+\frac{p_u \alpha_{\rm MRT}^2 \left|    \mathbf{g}_i^{H}\mathbf{g}_i \right|^2}{p_u \alpha_{\rm MRT}^2 \sum_{i'\neq i}^{n_r} \left|    \mathbf{g}_i^{H} \mathbf{g}_{i'}\right|^2+ \sigma^2_w}\right)  \right\}, 
	\end{aligned}
\end{equation}
 where $\mathbf{g}_i$ is the $i$-th column of matrix $\mathbf{H}_a^{H}$. The superscript $(m)$ is omitted for the purpose of simplification in the following derivation. To further derive the bound of the MRT precoding system, we adopt Jensen's inequality, specifically, 
\begin{equation}  
	\begin{aligned}
		&\mathcal{R}^{(m),i}_{({\rm MRT})}\\
		 &\!\geq \! \log_2\!\left(\!1+\! \left(\mathbb{E} \!\left \{\!\frac{\! p_u \!\alpha_{\rm MRT}^2 \!\sum_{i'\neq i}^{n_r}\! \left|    \mathbf{g}_i^{H} \mathbf{g}_{i'}\right|^2\!+\! \sigma^2_w } {p_u \alpha_{\rm MRT}^2 \left|    \mathbf{g}_i^{H}\mathbf{g}_i \right|^2} \!\right\}\!\right)^{-1}  \!\right)\!, 
	\end{aligned}
\end{equation}
where $\|[\mathbf{H}^{(m)}_a]_{i,:} \mathbf{g}_{i'} \|^2=n_s \hat{\sigma}_s^2 \sigma_{r,i}^2$, $\|[\mathbf{H}^{(m)}_a]_{i,:} \mathbf{g}_{i } \|^2=(n_s^2+n_s) \hat{\sigma}_s^2 \sigma_{r,i}^2$, and $\| \mathbf{V}  \|_{F}^2=\sum_{i=1}^{n_r} \sum_{j=1}^{n_s} \sigma_{r,i}^2 \sigma_{s,i}^2\approx n_s \hat{\sigma}_s^2 \sum_{i=1}^{n_r} \sigma_{r,i}^2$ .

Due to 
\begin{equation} \label{equ:mrt_inverse}
	\begin{aligned}
		&\mathbb{E}\left\{\frac{p_u \alpha_{\rm MRT}^2  \sum_{i \neq i'}^{n_r}\left|\mathbf{g}_{i}^{H} \mathbf{g}_{i'}\right|^{2}+\sigma^2_w}{p_u \alpha_{\rm MRT}^2  \left\|\mathbf{g}_{i}\right\|^{4}}\right\}\\
		&\!=\! \left(\! \sum_{i' \neq i}^{n_r}\! \mathbb{E}\!\left\{\!\left|\tilde{g}_{i'}\right|^{2}\!\right\}\! \right)\! \mathbb{E}\!\left\{\!\frac{1}{   \left\|\mathbf{g}_{i}\!\right\|^{2}}\!\right\}   \!+\!\mathbb{E}\! \left\{ \!\frac{\sigma^2_w} { p_u \alpha_{\rm MRT}^2 \left |\mathbf{g}_{i}^{H}  \mathbf{g}_{i} \right |^{2} } \!\right\}\!,
	\end{aligned}
\end{equation}
where $\tilde{g}_{i'}= \frac{\mathbf{g}_{i}^{H} \mathbf{g}_{i'} }{\|\mathbf{g}_{i}\|}$, and $\mathbf{g}_{i}$ is variable with zero mean and variance $ \beta_{i'}=\sigma_{r,i'}^2  \hat{\sigma}_{s}^2$, where $\hat{\sigma}_{s}^2=\frac{1}{n_s}\sum_{j=1}^{n_s} \sigma_{s,j}^2$, which is used to approximate the lower bound in derivation. 
\begin{equation} \label{equ:mrt_inverse_1}
	\mathbb{E}\left\{\frac{1}{   \left\|\mathbf{g}_{i}\right\|^{2}}\right\} =\frac{1}{\sum_{j=1}^{n_s} \sigma_{r,i}^2 \sigma_{s,j}^2 }=\frac{1}{(n_s-1) \hat{\sigma}_{s}^2 \sigma_{r,i}^2   }.
\end{equation}
\begin{equation} \label{equ:mrt_inverse_2}
	\mathbb{E}\left\{\frac{1}{   \left|\mathbf{g}_{i}^{H} \mathbf{g}_{i}\right|^{2}}\right\} =\frac{1}{(n_s-1) (n_s-2) \hat{\sigma}_{s}^4 \sigma_{r,i}^4 }.
\end{equation}

Substituting \eqref{equ:mrt_inverse_1} and \eqref{equ:mrt_inverse_2} into \eqref{equ:mrt_inverse}, we have

\begin{equation}
	\begin{aligned}
		&\mathbb{E}\left\{\frac{p_u \alpha_{\rm MRT}^2  \sum_{i \neq i'}^{n_r}\left|\mathbf{g}_{i}^{H} \mathbf{g}_{i'}\right|^{2}+\sigma^2_w}{p_u \alpha_{\rm MRT}^2  \left\|\mathbf{g}_{i}\right\|^{4}}\right\}\\
		&= \frac{ \hat{\sigma}_{s}^2  \sum_{i' \neq i}^{n_r}  \sigma_{r,i'}^2 }{(n_s-1) \hat{\sigma}_{s}^2 \sigma_{r,i}^2  }   +  \frac{\sigma^2_w}{ p_u \alpha_{\rm MRT}^2 (n_s-1) (n_s-2) \hat{\sigma}_{s}^4 \sigma_{r,i}^4 }\\ 
         &=\frac{ p_u  (n_s-2) \hat{\sigma}_{s}^2 \sigma_{r,i}^2 \sum_{i' \neq i}^{n_r}  \sigma_{r,i'}^2  + \sigma^2_w  n_s   \sum_{i=1}^{n_r} \sigma_{r,i}^2 } { p_u  (n_s-1) (n_s-2) \hat{\sigma}_{s}^2 \sigma_{r,i}^4 },  
	\end{aligned}
\end{equation}
where $\alpha_{\rm MRT} =   \sqrt{\frac{1}{n_s \hat{\sigma}_s^2 \sum_{i=1}^{n_r} \sigma_{r,i}^2}} $. Thus, the theoretical capacity bound is given by 
\begin{equation}
	\begin{aligned}
			&\tilde{\mathcal{R}}^{(m),i}_{({\rm MRT})}\! \\
			&\!\geq\! \log _{2}\!\!\left(\!\!1\!+ \!\frac{ p_u  (n_s-1) (n_s-2) \hat{\sigma}_{s}^2 \sigma_{r,i}^4}   { p_u  \!(n_s\!\!-\!2) \hat{\sigma}_{s}^2 \sigma_{r,i}^2 \!\sum_{i' \neq i}^{n_r} \! \sigma_{r,i'}^2 \! \!+\!\! \sigma^2_w   n_s  \! \!\sum_{i=1}^{n_r} \!\sigma_{r,i}^2 }     \!\!     \right) \\
		&\overset{\!n_s\!\! \gg 2}{=} \log _{2}\left(\!\!1\!+ \!\frac{ p_u  n_s    \hat{\sigma}_{s}^2  \sigma_{r,i}^4 }   { p_u    \hat{\sigma}_{s}^2 \sigma_{r,i}^2 \sum_{i' \neq i}^{n_r}  \sigma_{r,i'}^2  + \sigma^2_w    \sum_{i=1}^{n_r} \sigma_{r,i}^2  } \!\! \right). 
	\end{aligned}
\end{equation}
 
\subsection{ZF Precoding}
This precoding scheme intends at eliminating interference among different users by setting the precoding matrix as $\mathbf{V}=\alpha_{\rm ZF} \tilde{\mathbf{H}}_a^{H} \left(\tilde{\mathbf{H}}_a  \tilde{\mathbf{H}}_a^{H}\right)^{-1} $, where $\alpha_{\rm ZF}$ is the normalization factor to obey  the constraint $\mathbb{E}\{ \operatorname{Tr} ( \mathbf{V}    \mathbf{V} ^{H}  )\}=1$. In this case, it holds $\alpha_{\rm ZF} \tilde{\mathbf{H}}^{(m)}_a \mathbf{V}^{(m)}=\mathbf{I}_{n_r}$. We adopt the vector normalization method in \cite{6477575}, i.e., $\tilde{\mathbf{H}}_a^{H} \left(\tilde{\mathbf{H}}_a  \tilde{\mathbf{H}}_a^{H}\right)^{-1}=[\mathbf{f}_1, \ldots, \mathbf{f}_{n_r} ]$,  $\mathbf{V}_{:,i}=\frac{ \mathbf{f}_i}{\sqrt{n_r} \| \mathbf{f}_i \| }$. It should be noted that MRT and ZF adopt  different normalization methods to reduce noise impact for better exploitation of performance. Specifically, from analytical numerical results in \cite{6477575}, MRT with matrix normalization is better than vector normalization, and ZF with vector normalization is better than that with matrix normalization in achievable rate bounds. Thus,  the achievable rate is given by 
\begin{equation}  
	\begin{aligned}
		&\mathcal{R} ^{(m),i}_{({\rm ZF})} \!  =\! \mathbb{E} \left \{\!  \log_2\left(\! 1\! +\! \frac{p_u  \!  |[\tilde{\mathbf{H}}^{(m)}_a]_{i,:} [ {\mathbf{V}}^{(m)} ]^{-1}_{:,i}\|^2  }{p_u  | [\tilde{\mathbf{H}}^{(m)}_a]_{i,:}  \sum_{i'\neq i}^{n_r} [\tilde{\mathbf{H}}^{(m)}_a]^{-1}_{:,i'}  |^2  \! +\!  \sigma^2_w   } \! \right) \!  \right\} \! \\
		&=\mathbb{E} \left \{ \log_2\left(1\! +\! \frac{p_u   |[  \mathbf{H}^{(m)}_a]_{i,:} [ \mathbf{H}^{(m)}_a]^{-1}_{:,i}|^2  }{p_u \alpha_{\rm ZF}^2 | [ \mathbf{H}^{(m)}_a]_{i,:}  \sum_{i'\neq i}^{n_r}  [ \mathbf{H}^{(m)}_a]^{-1}_{:,i'}  |^2 \!  + \! \sigma^2_w}\! \right)  \! \right\} \! \\
		&=\mathbb{E} \left \{ \log_2\left(1+\frac{p_u  |[  \mathbf{H}^{(m)}_a]_{i,:} \frac{ \mathbf{f}_{i}}{\sqrt{n_r} \| \mathbf{f}_{i} \| }  |^2  }{p_u   | [ \mathbf{H}^{(m)}_a]_{i,:}  \sum_{i'\neq i}^{n_r}  \frac{ \mathbf{f}_{i'}}{\sqrt{n_r} \| \mathbf{f}_{i'} \| }    |^2  + \sigma^2_w}\right)  \right\} \\
		&= \mathbb{E} \left \{ \log_2\left(1+ {  \frac{p_u}{ {n_r} \sigma^2_w \| \mathbf{f}_i \|^2 }       } \right)  \right\}.  
	\end{aligned}
\end{equation}

We adopt the mathematical method in \cite{wong2008array} to analyze the theoretical capacity. Due to $\mathbf{f}_i=\tilde{\mathbf{H}}_a^{H} \left(\tilde{\mathbf{H}}_a  \tilde{\mathbf{H}}_a^{H}\right)^{-1} \mathbf{e}_{i}$, where $\mathbf{e}_{i}$ is a column vector that is $1$ at $i$-th entry and $0$ otherwise, the term $\frac{1}{   \| \mathbf{f}_i \|^2 }$ can be given as  
\begin{equation}
\begin{aligned}
	\beta_{i}&=\frac{1}{   \| \mathbf{f}_i \|^2 } =\frac{1}{   \| \tilde{\mathbf{H}}_a^{H} \left(\tilde{\mathbf{H}}_a  \tilde{\mathbf{H}}_a^{H}\right)^{-1} \mathbf{e}_{i} \|^2 }\\
	&= \frac{1}{ \mathbf{e}_{i}^{T} \left( {\mathbf{H}}_a   {\mathbf{H}}_a^{H}\right)^{-1} \mathbf{e}_{i} } =\frac{\operatorname{det} [\mathbf{H}_a \mathbf{H}_a^{H}] }{\operatorname{det} [\mathbf{H}_a^{(i)-} [\mathbf{H}_a^{(i)-}]^{H}] },
\end{aligned}
\end{equation}
where $\mathbf{H}_a^{(i)-}$ is the matrix of $\mathbf{H}_a$ deleting $i$-th row, and $\operatorname{det} $ denotes the determinant of a matrix. We have  \cite{wong2008array} 
\begin{equation}
	\begin{aligned}
		\operatorname{det} [\mathbf{H}_a \mathbf{H}_a^{H}] = \sum_{i=1}^{n_r} (-1)^{i-1} \mathbf{g}_1^{H} \mathbf{g}_i   \operatorname{det} [\mathbf{H}_a^{(1)-} [\mathbf{H}_a^{(i)-}]^{H}]. 
	\end{aligned}
\end{equation}

Thus,
\begin{equation}
	\begin{aligned}
		&\beta_{1}=\mathbf{g}_1^{H} \mathbf{g}_1 - \frac{ \sum_{i=2}^{n_r} (-1)^{i} \mathbf{g}_1^{H} \mathbf{g}_i   \operatorname{det} [\mathbf{H}_a^{(1)-} [\mathbf{H}_a^{(i)-}]^{H}]} {\operatorname{det} [\mathbf{H}_a^{(1)-} [\mathbf{H}_a^{(1)-}]^{H}] }\\
		&=\!\mathbf{g}_1^{H}\! \mathbf{g}_1 \! -\! \sum_{i=2}^{n_r}\! (\!-1\!)^{i}\! \mathbf{g}_1^{H}\! \mathbf{g}_i  \frac{\sum_{k=2}^{n_r} (-1)^{k-1} \mathbf{g}_k^{H} \mathbf{g}_1 \operatorname{det} [\mathbf{M}_k]  }{\!\sum_{j=2}^{n_r} (-1)^{i+j-1} \mathbf{g}_j^{H} \mathbf{g}_i \operatorname{det} [\mathbf{M}_j]\! }\\
		&=\mathbf{g}_1^{H} \mathbf{g}_1 -  \sum_{i=2}^{n_r}  \frac{\sum_{k=2}^{n_r} (-1)^{k-1} \mathbf{g}_k^{H} \mathbf{g}_1  \mathbf{g}_1^{H} \mathbf{g}_i  \operatorname{det} [\mathbf{M}_k]  }{\sum_{j=2}^{n_r} (-1)^{j-1} \mathbf{g}_j^{H} \mathbf{g}_i \operatorname{det} [\mathbf{M}_j] },
	\end{aligned}
\end{equation}
where $\mathbf{M}_k$ is the sub-block matrix of $\mathbf{H}_a^{(1)-} [\mathbf{H}_a^{(1)-}]^{H}$ with removal of the $k$-th row and the $i$-th column. Using the fact that each element in $\mathbf{H}_a$ is independent, and $\mathbb{E}\{[\mathbf{H}_a]_{i,j}^2\}=\sigma_{r,i}^2 \sigma_{s,j}^2$ under separable scattering environment, we have  $\mathbb{E}\{[\mathbf{g}_1 \mathbf{g}_1^H]\}=\operatorname{diag} [\sigma_{r,1}^2 \sigma_{s,1}^2,\ldots,\sigma_{r,1}^2 \sigma_{s,n_s}^2]$. Thus,  
\begin{equation}
	\begin{aligned}
		&\mathbb{E} \left\{ \frac{\sum_{k=2}^{n_r} (-1)^{k-1} \mathbf{g}_k^{H} \mathbf{g}_1  \mathbf{g}_1^{H} \mathbf{g}_i  \operatorname{det} [\mathbf{M}_k]  }{\sum_{j=2}^{n_r} (-1)^{j-1} \mathbf{g}_j^{H} \mathbf{g}_i \operatorname{det} [\mathbf{M}_j] } \right\}\\
		&=\mathbb{E} \left\{ \frac{\sum_{k=2}^{n_r} (-1)^{k-1} \mathbf{g}_k^{H}\mathbb{E}\{ \mathbf{g}_1  \mathbf{g}_1^{H} \} \mathbf{g}_i  \operatorname{det} [\mathbf{M}_k]  }{\sum_{j=2}^{n_r} (-1)^{j-1} \mathbf{g}_j^{H} \mathbf{g}_i \operatorname{det} [\mathbf{M}_j] } \right\}\\
		& \overset{(a)}{=} {\sigma}_{r,1}^2 \hat{\sigma}_s^2,  
	\end{aligned}
\end{equation}
where (a) simplifies the derivation  using average variances, i.e., $\mathbb{E}\{[\mathbf{g}_1 \mathbf{g}_1^H]\}= {\sigma}_{r,1}^2 \hat{\sigma}_s^2 \mathbf{I}_{n_s}$. As a result, 
\begin{equation}
	\begin{aligned}
		&\mathbb{E}\{\beta_{1}\}=\mathbb{E}\{\frac{1}{   \| \mathbf{f}_1 \|^2 }\}  =\mathbb{E}\{\mathbf{g}_1^{H} \mathbf{g}_1\} \\
		&\quad -  \mathbb{E}\left\{\sum_{i=2}^{n_r}   \frac{\sum_{k=2}^{n_r} (-1)^{k-1} \mathbf{g}_k^{H} \mathbf{g}_1  \mathbf{g}_1^{H} \mathbf{g}_i  \operatorname{det} [\mathbf{M}_k]  }{\sum_{j=2}^{n_r} (-1)^{j-1} \mathbf{g}_j^{H} \mathbf{g}_i \operatorname{det} [\mathbf{M}_j] } \right\} \\
		&\approx \sigma_{r,1}^2 \sum_{j=1}^{n_s} \sigma_{s,j}^2 -(n_r-1)  {\sigma}_{r,1}^2 \hat{\sigma}_s^2\\
		&= (n_s-n_r+1) \sigma_{r,1}^2 \hat{\sigma}_{s}^2.  
	\end{aligned}
\end{equation}

Based upon the above observation, we can derive the theoretical capacity as 
\begin{equation}  
	\begin{aligned}
		\tilde{\mathcal{R}} ^{(m),i}_{({\rm ZF})} &= \mathbb{E} \left \{ \log_2\left(1+ {  \frac{p_u}{ {n_r} \sigma^2_w \| \mathbf{f}_i \|^2 }       } \right)  \right\} \\
		&\approx \log_2\left(1+ {  \frac{p_u  }{ {n_r} \sigma^2_w  }   (n_s-n_r+1) \sigma_{r,i}^2 \hat{\sigma}_{s}^2       } \right). 
	\end{aligned}
\end{equation}
 
\subsection{NS-Based ZF Precoding} 
The ZF precoding scheme involves the inverse operation of the $n_r\times n_r$ matrix, a traditional approach is to compute the exact inverse of the matrix $\mathbf{Z}=\mathbf{H}_a \mathbf{H}_a^{H}$ in 
\begin{equation}
	\mathbf{V}=\alpha_{\rm ZF} \mathbf{H}_a^{-1}=\mathbf{H}_a^{H} \left(\mathbf{H}_a \mathbf{H}_a^{H}\right)^{-1} = \mathbf{H}_a^{H}\mathbf{Z}^{-1}.
\end{equation}

However, the large number of patch antennas  in MU-HMIMOS  communications imposes challenges to precoding design in practice. Specifically, the computation of matrix inversion would be excessively high as $n_r$ grows large, and even for the simple linear ZF precoding  that is still  impractical.  Thus, we employ NS expansion to replace the matrix inversion. Specifically, in separable scattering environment, 
\begin{equation}
\begin{aligned}
	\mathbf{Z}^{-1}&= \left(\mathbf{H}_a \mathbf{H}_a^{H}\right)^{-1} \\
	&= \left(\operatorname{diag} (\boldsymbol{\sigma}_r) \mathbf{W} \operatorname{diag} (\boldsymbol{\sigma}_s^2) \mathbf{W}^{H} \operatorname{diag} (\boldsymbol{\sigma}_r) \right)^{-1} \\
	&=   \operatorname{diag} (\boldsymbol{\sigma}_r) ^{-1}   \tilde{ \mathbf{W}}^{-1} \operatorname{diag} (\boldsymbol{\sigma}_r) ^{-1}  ,
\end{aligned}
\end{equation}
where $\tilde{ \mathbf{W}}= \mathbf{W} \operatorname{diag} (\boldsymbol{\sigma}_s^2) \mathbf{W}^{H}$, and $\boldsymbol{\sigma}_s^2$ is the element-wise square operation of $\boldsymbol{\sigma}_s$. $\operatorname{diag} (\boldsymbol{\sigma}_s^2)$ and	 $\operatorname{diag} (\boldsymbol{\sigma}_r)$ are diagonal matrix that are easily to compute the matrix inversion, thus the computational cost mainly lies in the computation of $\tilde{\mathbf{W}}^{-1}$. However, if Gram matrix $\tilde{ \mathbf{W}}$ is not a strongly diagonally dominant or even not dominant at all, the Neumann method cannot be applied in the computation of $\tilde{\mathbf{W}}^{-1}$ directly, which may result in the slow convergence or even divergence \cite{7248580}. Fortunately, in the special case of single ended correlation, (i.e., transmit correlation, or receive correlation, but not both are considered), the Neumann method works well as follows. 

In Neumann method, the Gram matrix $\tilde{ \mathbf{W}}$ is decomposed into its main diagonal matrix $\mathbf{D}$ and off-diagonal matrix $\mathbf{E}$ \cite{7248580}, i.e., $\tilde{ \mathbf{W}}=\mathbf{D}+\mathbf{E}$. The inverse of $\tilde{ \mathbf{W}}$ is given by 
\begin{equation}
	\tilde{ \mathbf{W}}^{-1}=\sum_{i=0}^{\infty} \left(-\mathbf{D}^{-1} \mathbf{E} \right)^{i} \mathbf{D}^{-1}.
\end{equation}

The inverse of $\tilde{\mathbf{W}}$ is approximated by a summation of powers of a matrix (matrix multiplications), which has a complexity order $\mathcal{O}\left(i n_r^3\right)$ ($i$ is the iteration number). Although the complexity order can be equal or higher (depending on iteration number $i$) than computing the exact inverse, matrix multiplications are preferable in hardware compared to the exact inversion \cite{6554990}.  In addition, increasing the iteration number $i$ brings a higher precision of the matrix inversion at a higher computational cost. In this paper, the iteration number $i$ is set to be $4$ for efficient computation at the low computational cost.


 \section{Numerical Results}\label{sec:simu}
In this section, we present computer simulation results of the downlink SE in the considered MU-HMIMOS system as well as the theoretical capacity. The single-sided correlation \cite{6297472} and three users are considered. All capacity curves were obtained after averaging over $800$ independent Monte Carlo channel realizations.  

\begin{figure}  
	\begin{center}
		\includegraphics[width=0.45\textwidth]{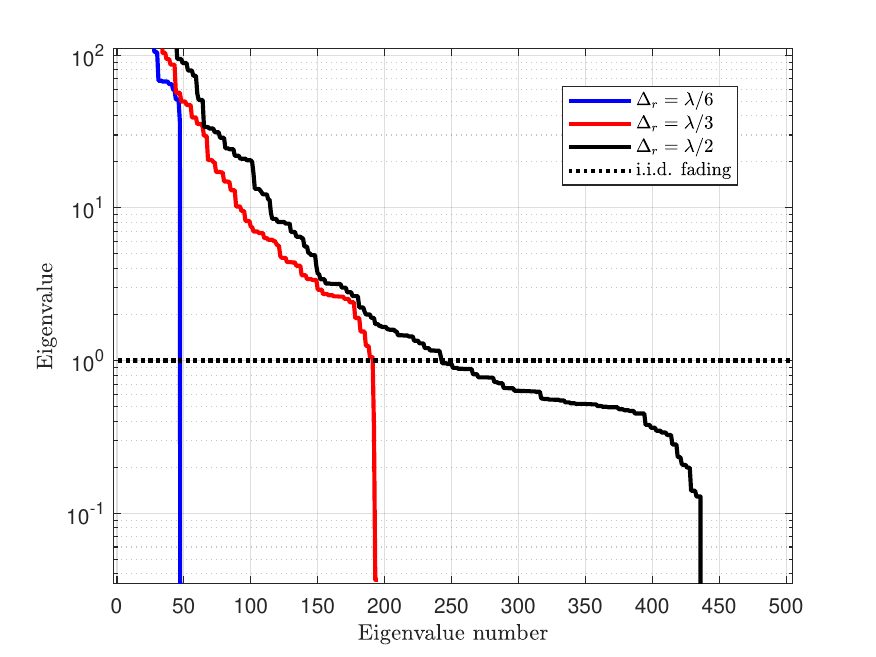}  \vspace{-2mm}
		\caption{The eigenvalues of $\mathbf{R}$ in decreasing order for MU-HMIMOS communication system with $N_s=900,N_r=576, \Delta_s=\lambda/3$ and $\Delta_r\in \{\lambda/6,\lambda/3,\lambda/2\}$.}
		\label{fig:EigvaluesCorNr576} \vspace{-6mm}
	\end{center}
\end{figure}

Fig$.$~\ref{fig:EigvaluesCorNr576} illustrates the eigenvalues of channel correlation matrix $\mathbf{R}_{\mathbf{H}}$ in decreasing order in a setup with $N_s=900,N_r=576, \Delta_s=\lambda/3$ for different spacing in received patch antennas. From the figure,  eigenvalues are large but non-identical initially, and then the eigenvalues quickly approach zero. This means the strengths of the coupling coefficients are not all equal even in isotropic propagation, which shows that the MU-HMIMOS channel exhibits spatial correlation. In addition, the smaller spacing among patch antennas, the more uneven the coupling coefficients and the steeper the eigenvalues decay, which implies stronger correlation. Specifically, the curve with $\Delta_r=\lambda/2$ decays much slower than that with $\Delta_r=\lambda/6$. The i.i.d. Rayleigh is also showed in black dot curve as reference. Normally, the spacing $\lambda/2$ is regarded as a threshold to maintain zero correlation among antennas, however, this is only an ideal assumption. Actually, the $\lambda/2$ still shows correlation in practice, thus, there is a difference between the case $\lambda/2$ and i.i.d. case.  We can see from the figure that none of the curves resembles the reference case, even the curve $\Delta_r=\lambda/2$ that is the closest one still has a major difference. These observations all prove that an EM channel in MU-HMIMOS systems should not adopt i.i.d. Rayleigh fading modeling. 

\begin{figure}  
	\begin{center}
		\includegraphics[width=0.45\textwidth]{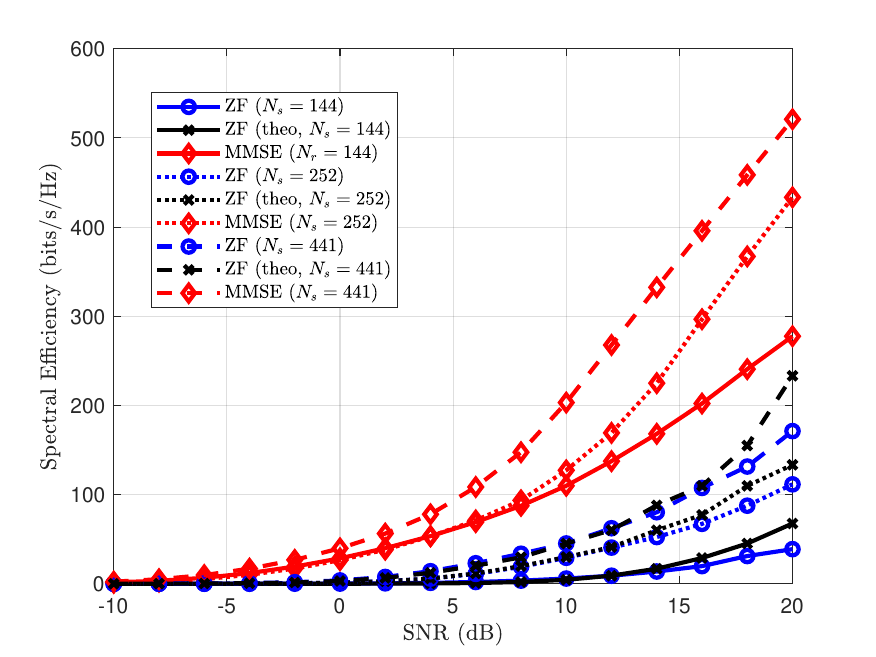}  \vspace{-2mm}
		\caption{SE of ZF precoding, theoretical ZF precoding and MMSE precoding schemes for different transmit patch antennas with $N_r=144, \Delta_s= \Delta_r=\lambda/3$.}
		\label{fig:Nr144s3r3_MMSE_ZF} \vspace{-6mm}
	\end{center}
\end{figure}

Fig$.$~\ref{fig:Nr144s3r3_MMSE_ZF} shows the SE of ZF precoding and theoretical ZF precoding schemes with $N_r=144, \Delta_s=\Delta_r=\lambda/3$ for different number of transmit patch antennas, and MMSE precoding scheme is set as benchmark curve. As shown in figure, there is very small gap between the ZF precoding and theoretical ZF schemes. Specifically, at lower SNR, the theoretical ZF perfectly predicts the ZF precoding scheme. Thus, the effectiveness of the presented theoretical capacity is proved. Naturally, the more transmit patch antennas bring more benefits in SE. This can be accounted for that transmit surface is larger with the increase of patch antennas given the fixed spacing. What's more, the gap between ZF precoding and  benchmark MMSE precoding is narrower with the increase of transmit patch antennas in high SNR region. This is mainly because the noise has little impact on ZF precoding when SNR goes high.  

\begin{figure}  
	\begin{center}
		\includegraphics[width=0.45\textwidth]{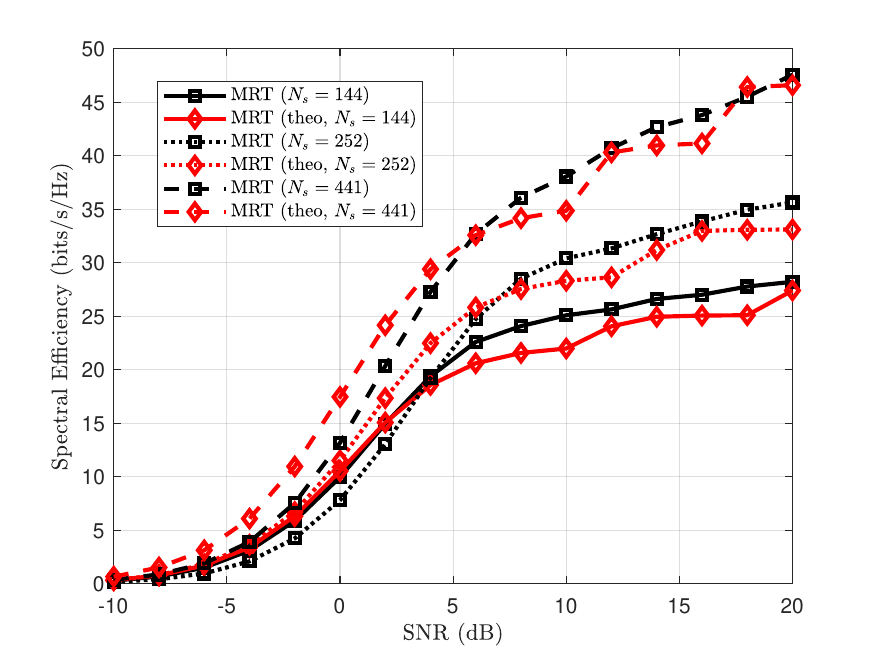}  \vspace{-2mm}
		\caption{SE of MRT precoding scheme  for different transmit patch antennas with $N_r=144, \Delta_s= \Delta_r=\lambda/3$.}
		\label{fig:Nr144s3r3_MRT} \vspace{-6mm}
	\end{center}
\end{figure}

The SE of MRT precoding and theoretical MRT precoding schemes with $N_r=144, \Delta_s=\Delta_r=\lambda/3$ for different number of transmit patch antennas are given in Fig.~\ref{fig:Nr144s3r3_MRT}. The similar conclusion drawn from Fig$.$~\ref{fig:Nr144s3r3_MMSE_ZF} can also be obtained in Fig.~\ref{fig:Nr144s3r3_MRT},  i.e., the theoretical MRT can predict MRT precoding scheme in all SNR regions. Shown in the figure, the curves for $N_s=144$ reach plateau at SNR=$10$ dB, this is because MRT shows advantages when the noise is dominant. With the increase of transmit patch antennas, not only the SE increases, but the point to reach plateau also moves further, and this phenomenon can be accounted for the larger transmit surface generated by the increase of patch antennas. 

\begin{figure}  
	\begin{center}
		\includegraphics[width=0.45\textwidth]{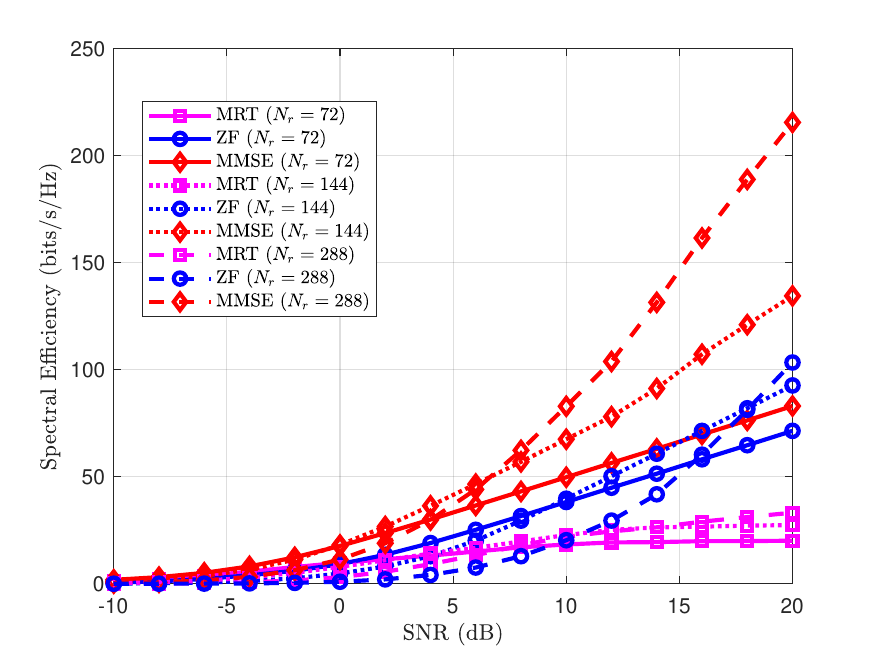}  \vspace{-2mm}
		\caption{SE of MRT, ZF and MMSE precoding schemes for different received patch antennas with $N_s=900,\Delta_s=\Delta_r=\lambda/6$.}
		\label{fig:Ns900s6r6} \vspace{-6mm}
	\end{center}
\end{figure}

The impact of received patch antennas on the SE of MRT, ZF and MMSE precoding schemes for $N_s=900, \Delta_s=\Delta_r=\lambda/6$ is given in Fig.~\ref{fig:Ns900s6r6}.   As shown in figure,  the MRT precoding is better than ZF precoding in the low SNR region, which is contrary to the case in the high SNR region. This is mainly due to the impact of noise, i.e., the noise is dominant in low SNR region, thus, MRT is better, while the noise impact is finite in the high SNR region, thus, ZF is better. As a benchmark, MMSE performs the best in the whole SNR region, and the gap between ZF and MMSE gradually decreases with the increase of SNR.  In addition, the more received patch antennas bring benefits in SE. As observed from figure, the case $N_r=288$ achieves the best performance compared with $N_r=72$ and $N_r=144$. This can be accounted for the larger received surface area enlarged by more received patch antennas under the fixed spacing. 

\begin{figure}  
	\begin{center}
		\includegraphics[width=0.45\textwidth]{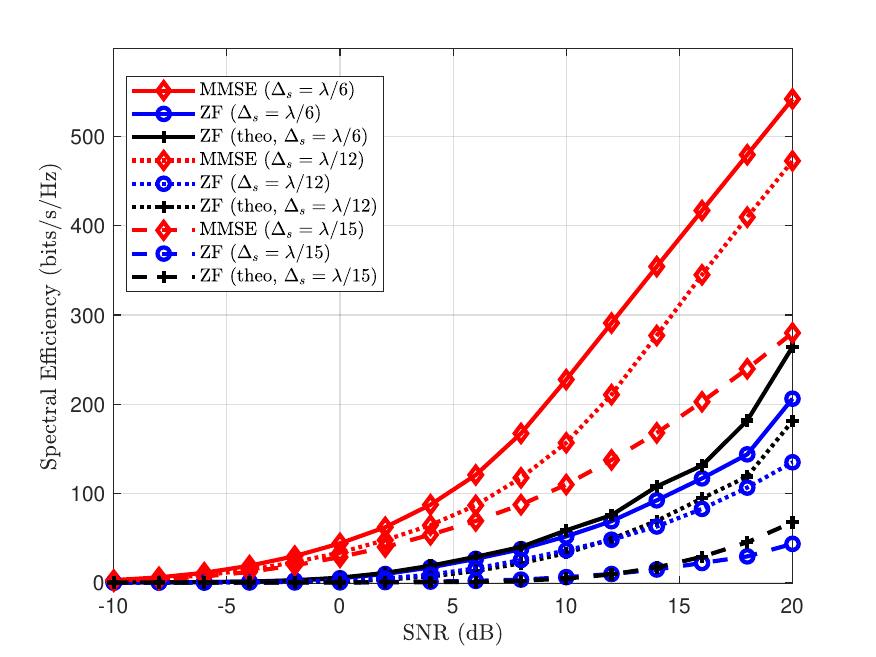}  \vspace{-2mm}
		\caption{SE of ZF and MMSE precoding schemes for different spacing with $N_s=3600,N_r=144,\Delta_r=\lambda/3$.}
		\label{fig:Ns3600Nr144r3_MMSE_ZF} \vspace{-6mm}
	\end{center}
\end{figure}

\begin{figure}  
	\begin{center}
		\includegraphics[width=0.45\textwidth]{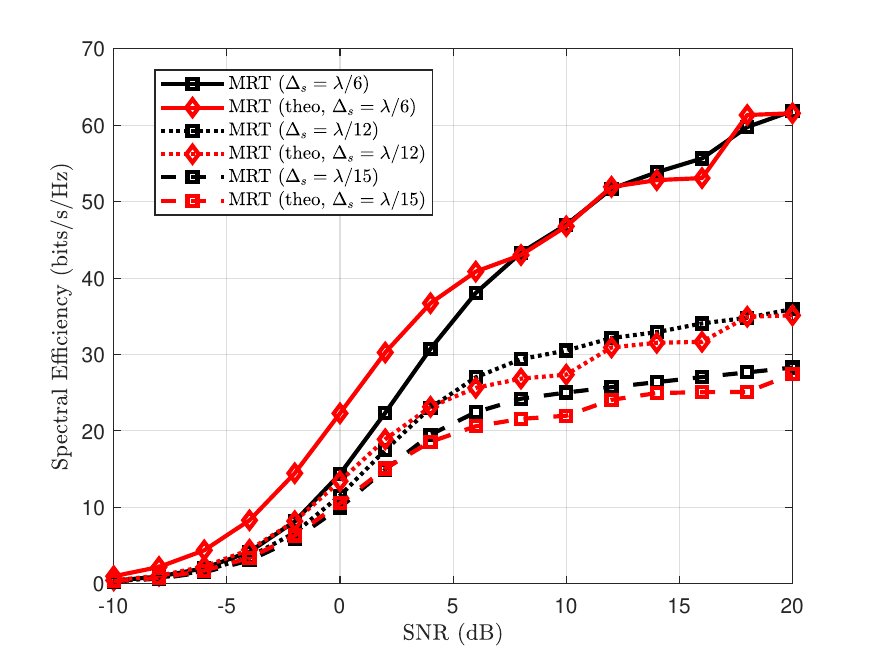}  \vspace{-2mm}
		\caption{SE of MRT  precoding schemes for different spacing with $N_s=3600,N_r=144,\Delta_r=\lambda/3$.}
		\label{fig:Ns3600Nr144r3_MRT} \vspace{-6mm}
	\end{center}
\end{figure}

The impact of transmit spacing on SE for MMSE, ZF, theoretical ZF, MRT and theoretical MRT precoding schemes are given in Fig$.$~\ref{fig:Ns3600Nr144r3_MMSE_ZF} and Fig.~\ref{fig:Ns3600Nr144r3_MRT}, respectively, under the settings of  $N_s=3600,N_r=144,\Delta_r=\lambda/3$. From the Fig$.$~\ref{fig:Ns3600Nr144r3_MMSE_ZF}, smaller spacing with fixed number of patch antennas has less surface area and induces more correlation among patch antennas, thus, the SE is worse. Specifically, the ZF precoding scheme  with $\Delta_s =\lambda/15$ has worse performance than  $\Delta_s =\lambda/6$. In other words, there is a large reduction in correlated channel with larger spacing. Normally, the spacing $\Delta_s=\Delta_r=\lambda/2$ is adopted in the uncorrelated channel assumption for SE analysis \cite{6457363}. In addition, the theoretical ZF with closer spacing predicts the ZF precoding well, thus, the presented theoretical ZF could perform well in highly correlated cases.  The similar observation is also obtained in Fig.~\ref{fig:Ns3600Nr144r3_MRT}. Specifically, the theoretical MRT with $\Delta_s=\lambda/15$ predicts that the performance of MRT is better than the case $\Delta_s=\lambda/6$ in the whole SNR region. Both figures show that less spacing has a lower SE and the proposed theoretical analyses perform well in highly correlated cases. 

\begin{figure}  
	\begin{center}
		\includegraphics[width=0.45\textwidth]{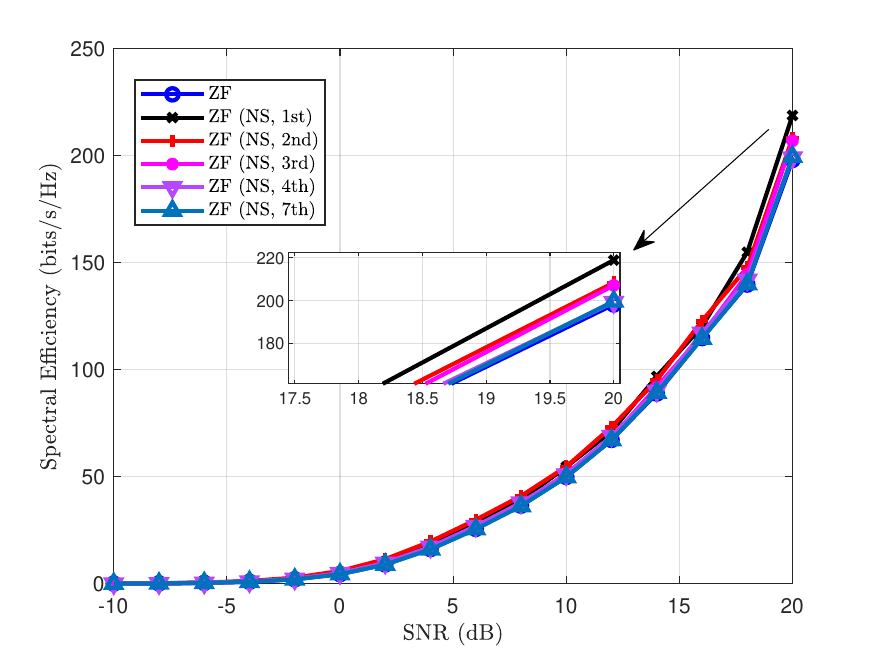}  \vspace{-2mm}
		\caption{Comparison of ZF precoding scheme and NS-based ZF precoding  for different iterations with $N_s=729,N_r=144,\Delta_s=\Delta_r=\lambda/3$.}
		\label{fig:NS_iter} \vspace{-6mm}
	\end{center}
\end{figure}

Fig.~\ref{fig:NS_iter} shows the impact of iteration on the SE for NS-based ZF precoding scheme with $N_s=729,N_r=144$ and $\Delta_s=\Delta_r=\lambda/3$. Shown in the figure, the ZF precoding employing NS with more iterations is closer to the ZF precoding scheme. Specifically, the case with $4$-th iteration almost coincides that with $7$-th iteration, and there is also just a narrow gap between $4$-th iteration and $3$-rd iteration. Thus, we normally adopt  $4$ iterations in practical simulations for achieving a balance between the computational cost and performance. The simulation result also proves the effectiveness of NS-based ZF precoding scheme since it achieves the similar performance with ZF while avoiding expensive matrix inversion operation.

\section{Conclusions}\label{sec:conclusion}
In this paper, we extended an EM-compliant channel model to MU-HMIMOS communication systems. The presented spatial channel is modeled in the wavenumber domain using the Fourier plane wave approximation, and it accounts for the inevitable mutual coupling induced by the close patch antenna spacing. We also analytically investigated the MRT and ZF precoding schemes, deriving the theoretical expressions for the achievable SE performance. Furthermore, a hardware efficient NS-based ZF precoding scheme was proposed to replace the involved matrix inversion, which renders the application of conventional ZF precoders impractical. Our simulation results verified that the more patch antennas with fixed spacing are present at the HMIMOS transmitter and receiver, the larger is the achievable SE performance. However, when the spacing among the antennas decreases, stronger signal correlations are generated, the SE is degraded. It was also demonstrated that the proposed NS-based ZF precoding scheme can achieve similar performance to conventional ZF at a lower hardware cost.


\bibliographystyle{IEEEbib}
\bibliography{strings}

\end{document}